\documentclass[letter]{aa} 

\usepackage{graphicx,txfonts,natbib,twoopt,breqn}
\usepackage{ulem}
 \usepackage[breaklinks=true]{hyperref} 
 \bibpunct{(}{)}{;}{a}{}{,}    
 \newcommandtwoopt{\citeads}[3][][]{\href{http://adsabs.harvard.edu/abs/#3}%
                                        {\citealp[#1][#2]{#3}}}
 \newcommandtwoopt{\citepads}[3][][]{\href{http://adsabs.harvard.edu/abs/#3}%
                                        {\citep[#1][#2]{#3}}}
 \newcommandtwoopt{\citetads}[3][][]{\href{http://adsabs.harvard.edu/abs/#3}%
                                        {\citet[#1][#2]{#3}}}
 \newcommandtwoopt{\citeyearads}[3][][]%
   {\href{http://adsabs.harvard.edu/abs/#3}{\citeyear[#1][#2]{#3}}}
\begin{document}

   \title{Detection of $J$-band photometric periodicity in the T8 dwarfs 2MASS J09393548-2448279 and EQ J1959-3338}
   \titlerunning{$J$-band variability in two T8 dwarfs}
   \author{P. A. Miles-P\'aez$^1$, S. Metchev$^{2,3}$, M. R. Zapatero Osorio$^1$ \and D. Mart\'in-Carrero$^4$
   }

   \institute{$^1$Centro de Astrobiolog\'ia, CSIC-INTA, Camino Bajo del Castillo s/n, 28692 Villanueva de la Ca\~nada, Madrid, Spain\\
              \email{pamiles@cab.inta-csic.es}\\
              $^2$Department of Physics and Astronomy, The University of Western Ontario, 1151 Richmond St, London, Ontario N6A 3K7, Canada\\
              $^3$Department of Physics and Astronomy, Institute for Earth and Space Exploration, The University of Western Ontario, 1151 Richmond St, London, Ontario N6A 3K7, Canada\\
            $^4$Departament d’Astronomia i Astrof\'isica, Universitat de Val\`encia, C. Dr. Moliner 50, 46100 Burjassot, Val\`encia, Spain
             }


 
  \abstract
   {}
   {We aim to study the near-infrared variability of the T8 dwarfs 2MASS J09393548-2448279 and EQ J1959-3338 by analyzing their $J$-band photometric signal, which can provide new insights into the atmospheric dynamics of cold brown dwarfs.}
   {We used FLAMINGOS-2 on the Gemini South telescope to perform $J$-band differential photometry continuously over 4 h for each target. The resulting light curves have a cadence of 20 s and a photometric uncertainty of 2-4 mmag.}  
   {We detect periodic variability in both T8 dwarfs, with an amplitude of $16.6\pm0.9$ mmag and a period of $1.364\pm0.012$ h for EQ J1959-3338, which we attribute to rotational modulation. For 2MASS J09393548-2448279, we observe an amplitude of $4.6\pm0.4$ mmag and a period of $1.733\pm0.040$ h, though this periodicity could represent a fraction of a longer period.}
   {With the detection of variability in 2MASS J09393548-2448279 and EQ J1959-3338, the number of known variable T8 dwarfs has doubled, making them prime candidates for infrared space-based monitoring and radio observations to investigate atmospheric dynamics and the influence of the magnetic field in very cool atmospheres.}
     

   \maketitle
%

\section{Introduction}

Photometric and spectroscopic monitoring of late M, L, and T dwarfs has revealed rotationally modulated variability, traditionally associated with patchy atmospheres containing clouds of silicates \citep[e.g., ][]{2009ApJ...701.1534A,2012ApJ...750..105R,2015ApJ...799..154M}. Different multiwavelength variability studies using space facilities such as the \textit{Hubble} Space Telescope (HST) or, more recently, the \textit{James Webb} Space Telescope (JWST) have revealed more complex atmospheric dynamics, including multiple cloud layers, disequilibrium chemistry processes, and rapid rotational weather patterns \citep[e.g., ][]{2012ApJ...760L..31B,2017Sci...357..683A,2024MNRAS.532.2207B}.

For cooler brown dwarfs at the T-to-Y transition, theory predicts a new sequence of sulfide, salt, and water clouds \citep{2012ApJ...756..172M}. These condensates are also supposed to drive rotationally-modulated photometric variability at infrared wavelengths, and, in particular, in the $J$ band, where the emergent dwarf's flux is expected to be more sensitive to the presence of condensates \citep{2014ApJ...789L..14M}. Similarly to warmer spectral types, other mechanisms such as chemical disequilibrum  \citep{2020A&A...643A..23T} and auroral emission \citep{2019RSPTA.37780398H} are also expected to lead to infrared variability at specific wavelength ranges. Thus, the identification of new photometrically variable dwarfs at the T-to-Y transition and their characterization at different wavelengths will allow us to study the interplay of these mechanisms in very cold atmospheres, ultimately informing models of giant exoplanets with similar temperatures.

However, late-T and Y dwarfs are themselves so intrinsically faint that they have remained relatively unexplored for variability to date as they emit most of their flux at wavelengths greater than $3\,\mu$m, whereas most variability searches have explored bluer wavelengths. For example, $J$-band variations have so far been reported only in the young T8 dwarf Ross 458C \citep{2019ApJ...875L..15M}, and marginally in one T7.5 \citep{2017MNRAS.466.4250L} and one Y0 dwarf \citep{2016ApJ...830..141L}.  In this letter, we present the photometric monitoring of two T8 dwarfs, 2MASS J09393548-2448279 and EQ J1959-3338, for which we find $J$-band periodic variability compatible with rotation.

\begin{figure*}
\centering
\includegraphics[width=0.93\textwidth]{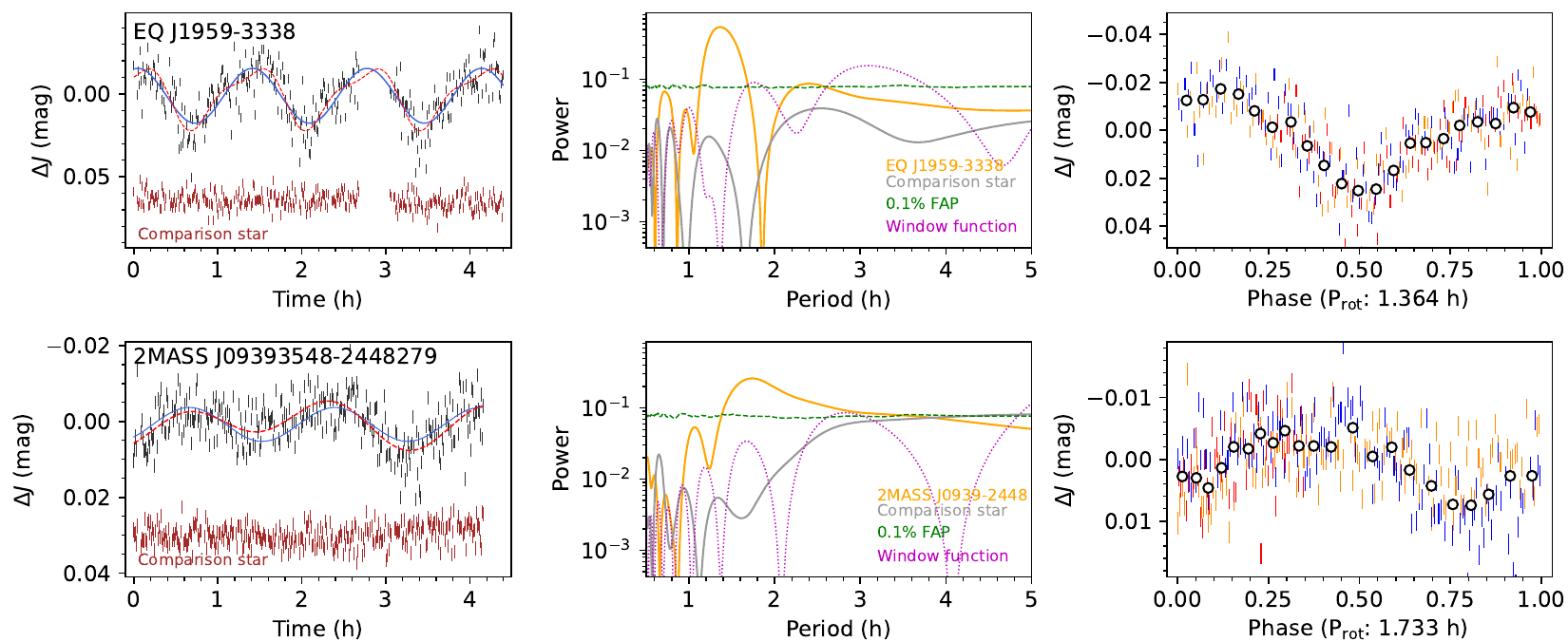}
\caption{{\sl Left}: $J$-band differential photometry of EQ J1959-3338 (top) and 2MASS J09393548-2448279 (bottom). Gray points represent 20 s photometric measurements, with symbol size indicating uncertainty. Comparison stars (brown symbols) are also shown. Data are vertically shifted for clarity. Best fits using a first-order truncated Fourier series (blue line) and a second-order series (red dashed line) are overlaid. {\sl Middle:} Lomb-Scargle periodogram of the left-panel time series. Targets (orange line) show strong periodicity at 1.4-1.9 h with 99.9\% confidence (green line), while comparison stars (gray line) show no significant periodicity. {\sl Right:} Phase-folded light curves of the targets, covering 2.9 (EQ J1959-3338) and 2.3 (2MASS J09393548-2448279) rotation cycles. Different colors (blue, orange, and red) represent separate cycles. Open circles show data binned every 15 points ($\approx$5 min).
}

\label{f1}%
\end{figure*}

\section{Targets}

EQ J1959-3338 ($J=16.87\pm0.15$ mag) has a trigonometric parallax of 85.3$\pm$2.2 mas \citep[11.7$\pm$0.3 pc, ][]{2019ApJS..240...19K} and a bolometric luminosity of $\log (L_{\rm bol}/L_\odot) = -5.558\pm0.021$ \citep{2024ApJ...973..107B}. Since there are no age constraints for EQ J1959-3338, \citet{2024ApJ...973..107B} used evolutionary models \citep{marley_2021_5063476} to derive a radius of 0.8-0.9 $R_{\rm jup}$ and an effective temperature of $T_{\rm eff} = 805^{+24}_{-43}$ K for a field age of 1-10 Gyr. This is in good agreement with the $T_{\rm eff}$ derived by modeling low-resolution near-infrared spectra of the object taken from the ground \citep{2021ApJ...921...95Z,2022ApJ...936...44Z} and from space \citep{2024ApJ...976...82T}. These physical parameters and evolutionary models point to a likely mass in the range 0.03-0.05 $M_\odot$. \citet{2015A&A...578A...1H} observed this target using adaptive optics in the near-infrared and found that EQ J1959-3338 appears to be single, with no detected companions down to $H=$20.7 mag at a 3$\sigma$ confidence level for separations $\ge$0\farcs5. No photometric monitoring for this target is available in the literature.

2MASS J09393548-2448279 ($J=15.98\pm0.11$ mag) has a trigonometric parallax of 187.3$\pm$4.6 mas \citep[5.34$\pm$0.13 pc, ][]{2019ApJS..240...19K} and a bolometric luminosity of $\log (L_{\rm bol}/L_\odot) = -5.768^{+0.066}_{-0.063}$ \citep{2021ApJ...921...95Z}. The modeling of its near-infrared spectra by \citet{2017ApJ...848...83L} and \citet{2022ApJ...936...44Z}  points to a cooler $T_{\rm eff}$ (611-632 K) and a radius slightly larger (1.01-1.22 $R_{\rm jup}$) than those of EQ J1959-3338. It has been suggested that 2MASS J09393548-2448279 is likely an unresolved equal-mass binary, with a mass range of 0.01-0.05 $M_{\odot}$ between the ages of 0.4 and 12 Gyr  \citep{2008ApJ...689L..53B}. \citet{2017ApJ...848...83L} report that the source remains unresolved down to 0\farcs07. \citet{2014ApJ...793...75R} obtained $J$-band differential photometry over 5.92 h using the WIRC instrument on the Du Pont 2.5 m telescope and did not find any variability down to 0.016 mag. \citet{2013AJ....145...71K} also monitored this target in the $K$ band over 2 h and reported a marginal detection of photometric variability at the level of 0.3 mag, but could not derive any rotation period.

According to the literature \citep{2022ApJ...936...44Z}, both dwarfs have an iron abundance compatible with solar levels, and their spectra do not significantly deviate from the sequence of known field late-T dwarfs \citep{2021ApJ...921...95Z}. We investigated whether our targets could be members of any known young moving groups \citep[e.g.,][]{2004ARA&A..42..685Z,2018ApJ...856...23G}. To this end, we calculated their UVW motions by combining trigonometric parallax and proper motions with the radial velocities reported in \citet{2021ApJ...921...95Z}: $-47\pm207$ km\,s$^{-1}$ for EQ J1959-3338 and $-84^{+243}_{-238}$ km\,s$^{-1}$ for 2MASS J09393548-2448279. Given the large uncertainties, neither target appears to belong to any young moving group for any plausible radial velocity within this range. The Hyades \citep[$\approx$600-800 Myr, ][]{2015ApJ...807...24B} is the oldest group we considered, suggesting that the kinematic age of our targets is likely greater than 600-800 Myr.

\section{Observations and data analysis \label{obs}}

We observed EQ J1959-3338 and 2MASS J09393548-2448279 using the $J$-band filter on FLAMINGOS-2 \citep{2004SPIE.5492.1196E} at Gemini South, collecting nearly 4 hours of imaging per target in staring mode, with individual exposure times of 20 s. To facilitate sky subtraction, a five-point dither pattern was applied every 20 minutes. The observations took place on May 1 and May 10, 2019, under clear skies, with seeing between 0\farcs6 and 1\farcs2. This strategy yielded 340 and 352 images per target. Standard reduction steps--including sky subtraction, flat-fielding, and alignment--were performed using IRAF \citep[][]{1986SPIE..627..733T}.

Differential photometry was applied to construct the light curves by selecting bright stars with stable flux as comparison, and by using an aperture of 1.5$\times$ the average Full-Width at Half Maximum (FWHM) computed for each image. This aperture minimized the sky noise contribution in our targets. The final light curves achieve median precisions of 2 mmag for 2MASS J09393548-2448279 and 4 mmag for EQ J1959-3338, and are shown in Fig. \ref{f1} (left). Both targets exhibit clear photometric modulations, unlike their stable comparison stars. We investigated potential systematic effects due to the object’s centroid position, airmass, or FWHM using the Pearson’s {\it r} correlation coefficient and find no significant trends. The full observational details and systematics analysis are provided in Appendices \ref{A1} and \ref{A2}.

\begin{figure}
\centering
\includegraphics[width=0.4\textwidth]{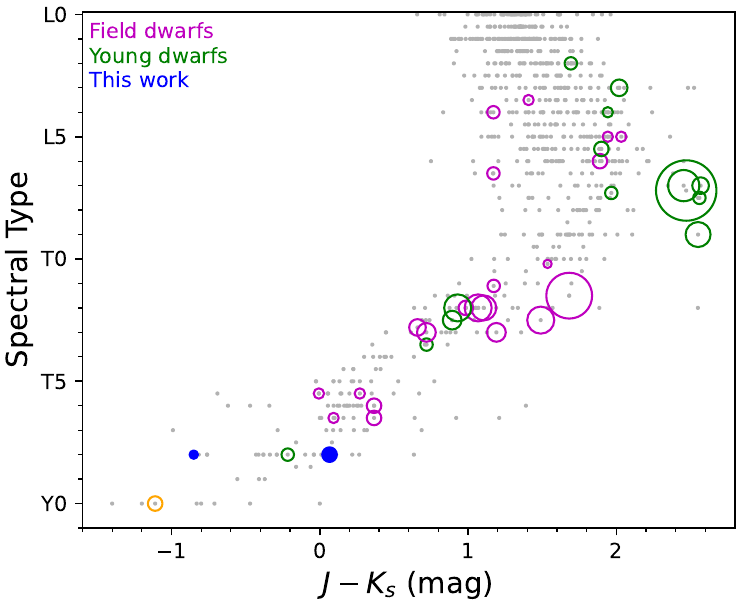}
\caption{Spectral type as a function of $J-K_s$ color for known L, T, and Y dwarfs from the compilation presented in \citet[][gray symbols]{best_2024_13993077}. Objects with a detection of $J$-band photometric variability are plotted in green for those thought to be young and magenta for those considered to have field ages. Their symbol size is proportional to the measured peak-to-peak amplitude. EQ J1959-3338 and 2MASS J09393548-2448279 are plotted with blue symbols. The open orange circle denotes the upper limit detection of the Y0 dwarf WISEP J173835.52+273258.9 \citep{2016ApJ...824....2L}. The different detections of $J$-band variability are taken from \citet{2008MNRAS.386.2009C,2009ApJ...701.1534A,2014ApJ...793...75R,2014ApJ...797..120R,2015ApJ...812..163B,2016ApJ...829L..32L,2016arXiv160903586C,2018AJ....155...95B,2019ApJ...875L..15M,2019MNRAS.483..480V,2019ApJ...883..181M,2019A&A...629A.145E,2020AJ....159..125L,2020ApJ...893L..30B,2024MNRAS.527.6624L}}
\label{f3}%
\end{figure}

\begin{figure}
\centering
\includegraphics[width=0.41\textwidth]{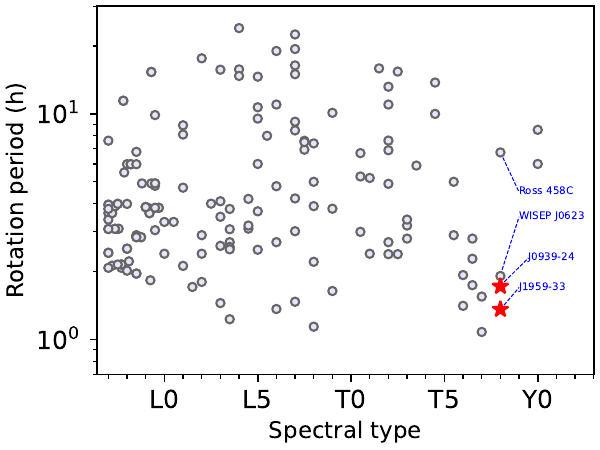}
\caption{Rotation period as a function of spectral type for all periodically variable M7-Y0 dwarfs known as of this writing. EQ J1959-3338 and 2MASS J09393548-2448279 are indicated by a red star. The position of the other two periodically variable T8 dwarfs (WISE J062309.94-045624.6 and Ross 458C) is also indicated. Literature values are taken from  \citet{2021AJ....161..224T,2022ApJ...924...68V,2023MNRAS.521..952M,2023ApJ...951L..43R}}
\label{f2}%
\end{figure}

\section{Search for periodicity\label{s5}}

We searched for periodic variability in the light curves of EQ J1959-3338 and 2MASS J09393548-2448279 using a Lomb-Scargle (LS) periodogram \citep[][Fig. \ref{f1}, middle]{1976Ap&SS..39..447L,1982ApJ...263..835S}. Each periodogram sampled $10^4$ frequencies corresponding to periods between 0.5-40 hours, a range typically expected for ultra-cool dwarfs \citep{2021AJ....161..224T}. A 0.1\% false-alarm probability (FAP) threshold was determined from $10^5$ randomizations as in \citet[][]{2017MNRAS.472.2297M}. Strong peaks were found close to  $\approx$1.40 h for EQ J1959-3338 and  $\approx$1.70 h for 2MASS J09393548-2448279, both well above the 0.1\% FAP and uncorrelated with window function artifacts. No significant periodicity was found in the comparison stars, confirming that the observed variability is intrinsic to the targets, likely due to rotation.

To estimate the variability properties, we fitted a sinusoidal model to the light curves using a Markov chain Monte Carlo (MCMC) approach with {\tt emcee} \citep{2010CAMCS...5...65G,2013PASP..125..306F}. The best-fit solutions yielded rotation periods and variability amplitudes of 1.364$\pm$0.012 h and 16.6$\pm$0.9 mmag for EQ J1959-3338, and 1.733$\pm$0.040 h and 4.6$\pm$0.4 mmag for 2MASS J09393548-2448279. These models are shown as a blue solid curve in Fig. \ref{f1} (left). The phase-folded light curves corresponding to these rotation periods are shown in the right panels of Fig. \ref{f1}.

While EQ J1959-3338 displays a well-defined periodic signal, the light curve of 2MASS J09393548-2448279 shows nonidentical minima in consecutive cycles, suggesting either an evolving atmospheric pattern, as seen in some L and early-T dwarfs \citep[e.g., ][]{2009ApJ...701.1534A,2017MNRAS.472.2297M}, or a longer true period. To explore deviations from a simple sine wave, we fitted a second-order truncated Fourier series, which better captures changes in amplitude by including an additional harmonic component (dashed red line in the left panels of Fig. \ref{f1}). However, this model provided rotation periods consistent with the sinusoidal fit, and the Bayesian information criterion (BIC) slightly favored the simpler sine model ($\Delta BIC$=3-5).

We further tested Gaussian processes (GP) by fitting our data to a sine function combined with a Mat\'ern-3/2 kernel \citep[e.g.,][]{2021A&A...651L...7M} and to the Rotation kernel \citep{2017AJ....154..220F}--which is commonly used for highly variable stars and brown dwarfs--to account for correlated noise \citep[e.g., ][]{2017MNRAS.466.4250L,2022ApJ...924...68V}. These advanced models also returned rotation periods consistent with our sinusoidal fits. Therefore, we adopted the sinusoidal model as our primary period estimate for EQ J1959-3338 and provisionally for 2MASS J09393548-2448279, which may require additional data to fully constrain its period. We note that period and amplitude uncertainties are purely statistical and do not include systematic errors arising from the sinusoidal assumption, which appear to be small based on tests with more complex models.

\section{Discussion and conclusion}

The observed $J$-band variability in Fig. \ref{f1} shows that the atmospheric heterogeneities seen in warmer spectral types persist even in very cold T dwarfs. In earlier-type brown dwarfs, rotational modulation is often attributed to patchy clouds or temperature variations \citep{2008MNRAS.391L..88L,2012ApJ...750..105R,2017ApJ...840...83M}. As these atmospheric  inhomogeneities rotate in and out of view, they modulate the observed flux, leading to periodic variability. To put our findings in context, we show in Fig. \ref{f3} the sample of young (green) and mature (pink) L, T, and Y dwarfs reported to exhibit $J$-band variability. The amplitudes of variability of EQ J1959-3338 and 2MASS J09393548-2448279 are similar to those seen in earlier field L and T dwarfs, with the exception of those at the L/T transition, which typically exhibit larger amplitudes--likely due to the disruption processes of silicate clouds taking place. Given the low $T_{\rm eff}$ of EQ J1959-3338 and 2MASS J09393548-2448279, their variability at 1.2 $\mu$m is likely linked to high-altitude condensate clouds composed of Na$_2$S or KCl as well as the likely presence of haze layers, which are predicted to appear in the cool atmospheres of late-T and Y dwarfs  \citep{2012ApJ...756..172M,2017ApJ...848...83L} and even giant exoplanets \citep{2024PSJ.....5..186H}. Hot spots \citep{2014ApJ...789L..14M} resulting, for example, from auroral activity \citep{2019RSPTA.37780398H,2024ApJ...966...58P,2024Natur.628..511F} could also explain some of the observed variability in our targets. \citet{2025ApJ...981L..22M} recently presented JWST spectroscopic time series for the T2 dwarf SIMP J013656.5+093347.3 and showed that the relative contribution of different emission mechanisms to the observed variability depends on the observation wavelength. Thus, identifying photometrically-variable dwarfs at later spectral types, such as EQ J1959-3338 and 2MASS J09393548-2448279, is crucial to study the relative contribution of different physical processes to the atmospheric dynamics and observed flux of objects at the colder T-to-Y transition.

Figure \ref{f2} shows the rotation periods currently known for all dwarfs with spectral types later than M7. Our targets occupy a sparsely populated region of parameter space, owing to the intrinsic faintness of these spectral types in the optical and infrared bands typically used to search for photometric variability in very low-mass stars and brown dwarfs. In fact, the periodicity of three out of the four T8-Y0 dwarfs shown in Fig. \ref{f2} was detected via infrared observations with the Spitzer Space Telescope and HST. The fourth object, the T8 dwarf WISE J062309.94-045624.6, had its rotation period identified through radio observations\citep[$P_{\rm rot} = 1.912\pm0.005$ h, ][]{2023ApJ...951L..43R}, and currently lacks variability studies at other wavelengths. Thus, EQ J1959-3338 and 2MASS J09393548-2448279 are only the third and fourth T8 dwarfs reported to exhibit variability. 

\begin{figure}
\centering
\includegraphics[width=0.4\textwidth]{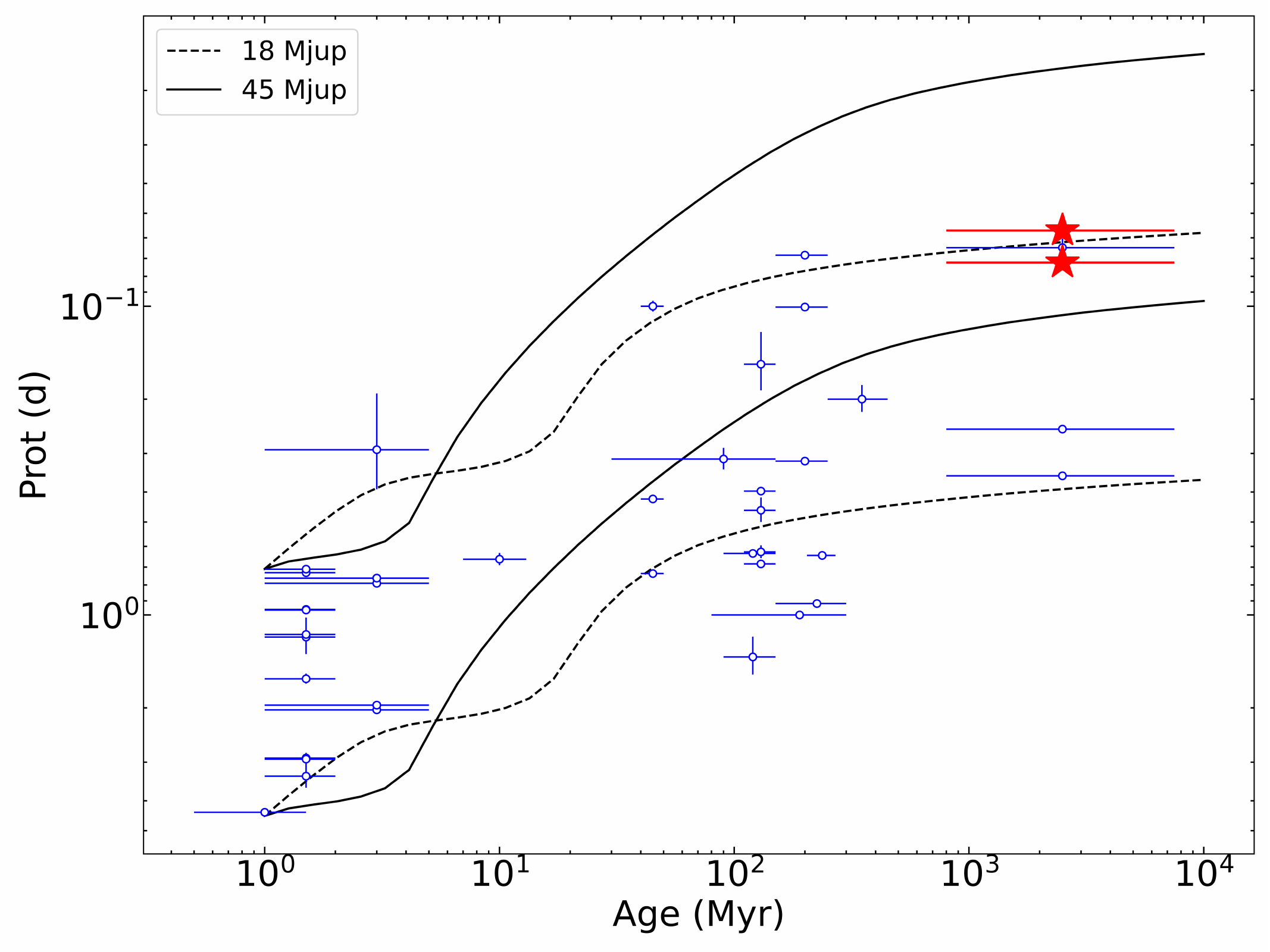}
\caption{Angular momentum evolution of 18-45 $M_{\rm jup}$ brown dwarfs. Our targets (red stars) and literature data (blue circles, see text) are shown. The solid and dashed lines represent angular momentum conservation models for two masses, using the radius evolution from  \citet{2023A&A...671A.119C}. The models start from the extreme rotation periods observed in the youngest objects.}
\label{f4}%
\end{figure}
 
The rotation periods of EQ J1959-3338 and 2MASS J09393548-2448279 are among the shortest observed for dwarfs of comparable mass. Figure \ref{f4} shows the rotation periods as a function of age for the mass range 18-45 $M_{\rm jup}$ \citep{cody10, crossfield14, scholz18, 2021AJ....161..224T, 2022ApJ...924...68V, hsu24,2024MNRAS.527.6624L}, alongside models of angular momentum conservation computed following  \citet{2006ApJ...647.1405Z} and using the substellar evolutionary tracks of \citet{2023A&A...671A.119C}. Despite the scatter at a given age, the observations suggest that the spin of these brown dwarfs increases over time, implying that any angular momentum loss must be moderate or small. This contrasts with solar-type stars, which become slow rotators by the age of 600--800 Myr \citep{1997ApJ...475..604S}. 

Different studies also show that rapidly-rotating T dwarfs may exhibit detectable radio emission by processes related to the electron cyclotron maser instability \citep{2015ApJ...808..189W,2017ApJ...846...75P,2019MNRAS.487.1994K}, including the late-T radio-burst emitter WISEP J062309.94-045624.6. The detection of photometric variability and fast rotation in EQ J1959-3338 and 2MASS J09393548-2448279 makes them excellent targets for dedicated deep radio surveys aimed at searching for radio-aurorae. Such observations can help determine how common these emissions are and their dependence on rotation, magnetic field strength, and atmospheric conditions.

\begin{acknowledgements}
We thank the anonymous referee for the thoughtful and supportive report. This work made use of the grant RYC2021-031173-I funded by MCIN/AEI/ 10.13039/501100011033 and by the ``European Union NextGenerationEU/PRTR'' as well as by the grant MDM-2017-0737. We acknowledge the support of the Canadian Space Agency through the Flights and Fieldwork for the Advancement of Science and Technology (FAST) program (reference No.\ 19FAWESB40). This work has benefited from The UltracoolSheet at http://bit.ly/UltracoolSheet, maintained by Will Best, Trent Dupuy, Michael Liu, Aniket Sanghi, Rob Siverd, and Zhoujian Zhang.
\end{acknowledgements}

%
%

\bibliographystyle{aa} 
\bibliography{biblio.bib} 

\begin{appendix}

\section{Observations \label{A1}}

We used the $J$-band filter and FLAMINGOS-2 \citep{2004SPIE.5492.1196E} on the Gemini South telescope to monitor each target for nearly 4 h in service mode. FLAMINGOS-2 has a 2048$\times$2048-pixel HAWAII-2 detector for the 0.9 to 2.5 $\mu$m range. Its pixel scale is 0\farcs18, which yields a circular field-of-view with a diameter of 6\farcm2. Observations were carried out on May 1, 2019, for 2MASS J09393548-2448279 and on May 10, 2019, for EQ J1959-3338 with air mass ranges of 1.01-1.85 and 2.05-1.07, respectively. The sky was clear and the seeing was typically  0\farcs6-1\farcs2 in both epochs.

We designed one-hour-long observing blocks (OBs) to collect imaging data in staring mode with individual exposure times of 20 s. The stare position was alternated with a five-point dither pattern every 20 min for sky subtraction with $x$- and $y$-offsets of 10$\arcsec$. Four OBs were obtained consecutively for each target, yielding a total of 340 and 352 images for EQ J1959-3338 and  2MASS J09393548-2448279, respectively. Observations at the end of the third OB of EQ J1959-3338 were interrupted for approximately 22 minutes due to a tracking issue but later resumed normally, creating a gap in the object's light curve (Fig. \ref{f1}, left). All data frames were sky-subtracted, flat-fielded using coeval dome flat images, and aligned using standard data reduction recipes and the Image Reduction and Analysis Facility \citep[IRAF;][]{1986SPIE..627..733T}.

\section{Data analysis \label{A2}}

We carried out differential photometry to obtain the light curves of EQ J1959-3338 and 2MASS J09393548-2448279 following \citet{2017MNRAS.472.2297M,2023MNRAS.521..952M}. After identifying all sources with signal-to-noise ratio (S/N) greater than 30 in the field of view of each target, we performed circular aperture photometry on these by using an aperture of 1.5$\times$ the average FWHM computed for each image. This aperture minimized the sky noise contribution in our targets. The photometry was sky-subtracted by computing the median value of the sky in an annulus with inner radius 3.5$\times$FWHM and a width of 2$\times$FWHM. Lastly, the differential light curves of our targets were built by dividing their fluxes by the sum of those in a set of 5 comparison stars, achieving a median photometric precision of 2 mmag for 2MASS J09393548-2448279 and 4 mmag for EQ J1959-3338. The comparison stars were chosen among those stars brighter ($\approx0.5-2$ mag) than our targets, but far from the nonlinear regime of the detector. The light curves of the comparison stars show the smallest standard deviation among all the stars in their fields of view. We visually inspected these light curves for variability to ensure that any detection of variability in EQ J1959-3338 and 2MASS J09393548-2448279 was not spuriously created. 

\begin{figure}
\centering
\includegraphics[width=0.5\textwidth]{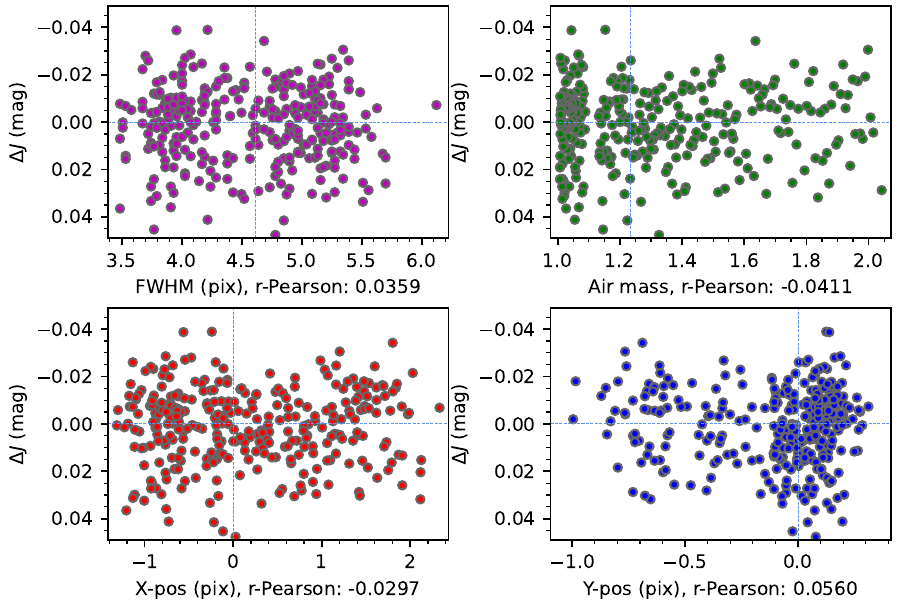}
\caption{Differential photometry of EQ J1959-3338 as a function of air mass, FWHM, and X and Y centroid position. The dashed, light blue lines indicate the median value of each variable. The Pearson's $r$ coefficient of linear correlation is also indicated for each pair.}
\label{app1}%
\end{figure}

\begin{figure}
\centering
\includegraphics[width=0.5\textwidth]{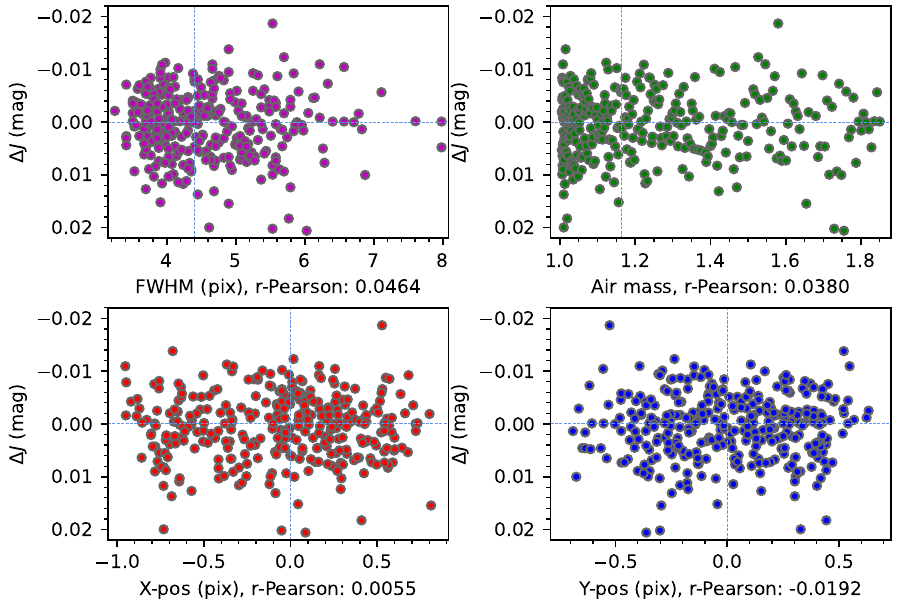}
\caption{Differential photometry of 2MASS J09393548-2448279 as a function of air mass, FWHM, and X and Y centroid position. The dashed, light blue lines indicate the median value of each variable. The Pearson's $r$ coefficient of linear correlation is also indicated for each pair.}
\label{app2}%
\end{figure}

The light curves of our targets are shown in Fig. \ref{f1} (left), in which we also plot the light curve of one of the comparison stars. There is an obvious modulation in the light curves of EQ J1959-3338 and 2MASS J09393548-2448279 compared to the relatively flat light curves of their comparison stars. In Figs. \ref{app1} and \ref{app2}, we plot the differential photometry of our targets as a function of their pixel position in both the x- and y-axes, the airmass, and the FWHM to search for the presence of residual systematics that can mimic photometric variability. We used the Pearson's {\it r} correlation coefficient to test for linear correlation attributable to any residual systematic. However, as we find no obvious correlation, we note that residual systematics are unlikely to contribute significantly to the observed photometric variability in the left panels of Fig. \ref{f1}.

\end{appendix}

\end{document}